\documentclass[preprint,prb,superscriptaddress,showpacs,preprintnumbers]{revtex4}
\usepackage{amsmath,dcolumn,graphicx,ulem}

\begin{document}

\title{Limitations on the extent of off-center displacements in TbMnO$_3$ from
EXAFS measurements}

\author{F. Bridges} \affiliation{Physics Department, University of California,
Santa Cruz, California 95064, USA}
\author{C. Downs } \affiliation{Physics Department, University of California,
Santa Cruz, California 95064, USA}
\author{T. O'Brien} \affiliation{Physics Department, University of California,
Santa Cruz, California 95064, USA},\affiliation{Los Alamos National 
Laboratory, Los Alamos, NM 87545}
\author{Il-K Jeong} \affiliation{Los Alamos National Laboratory, Los Alamos, NM 87545}
\author{T. Kimura} \affiliation{Los Alamos National Laboratory, Los Alamos, NM 87545}, \altaffiliation[Present address:]{Division of Materials Physics, Graduate School of Engineering Science, Osaka University, 1-3 Machikaneyama, Toyonaka, Osaka 560-8531, JAPAN}

\date{\today}

\begin{abstract}

We present EXAFS data at the Mn K and Tb L$_3$ edges that provide upper limits
on the possible displacements of any atoms in TbMnO$_3$. The displacements must
be less than 0.005-0.01\AA{} for all atoms which eliminates the possibility of
moderate distortions (0.02\AA) with a small {\it c}-axis component, but for
which the displacements in the {\it ab} plane average to zero. Assuming the
polarization arises from a displacement of the O2 atoms along the {\it c}-axis,
the measured polarization then leads to an O2 displacement that is at least
6X10$^{-4}$\AA, well below our experimental limit. Thus a combination of the
EXAFS and the measured electrical polarization indicate that the atomic
displacements likely lie in the range 6X10$^{-4}$ - 5X10$^{-3}$\AA.

\end{abstract}


\maketitle

\section{Introduction}

Multi-ferroics have been an area of intense investigation in recent years. In
such systems two or more orderings occur; in the case of TbMnO3 which we
discuss here, both antiferromagnetism and ferroelectricity co-exist at low T
and large magnetoelectric effects are observed\cite{Kimura03,Kimura05}. For
ferroelectricity to occur, inversion symmetry must be broken and there should
be displacements of some of the atoms in each unit cell to form an electric
dipole moment per cell; such a transition is called a displacive transition.
Alternatively inversion symmetry might already be broken locally with some
atoms displaced slightly off-center - perhaps at the Neal temperature T$_N$.
Then at temperatures above the ferroelectric transition temperature T$_c$, the
off-center displacement orientation is random, no net polarization occurs, and
on average, the center of positive charge is not displaced relative to the
center of negative charge. For such systems the ferroelectric transition is
called an order/disorder transition; here there is no change in the magnitude
of the local disorder at the nearest neighbor level, only an ordering of the
off-center displacements directions to form a net dipole moment.

To date the distortions that lead to ferroelectricity have not been identified
in TbMnO$_3$ although Kimura {\it etal}\cite{Kimura03} suggest it is likely the
O atoms that are displaced. Some studies suggest that there is a variation in
the buckling of the Mn-O-Mn linkage that is correlated with the
antiferromagnetic coupling.\cite{Aliouane06}  In that model, there will be O
displacements in different directions - only when the antiferromagnetic order
becomes commensurate with the lattice does it lead to a net displacement along
some axis and to ferroelectricity. 


The net magnitude of the displacement must be small based on the measured {\it
c}-axis ferroelectric polarization,\cite{Kimura03} P $\sim$ 800 $\mu$C/m$^2$.
This value of P corresponds to a dipole moment in each unit cell (volume $\sim$
229 \AA$^3$) of approximately 0.01 {\it e}\AA {} (1 {\it e}\AA {} is a charge
of 1 electron displaced 1 \AA).  Thus if there is one electron charge displaced
along the {\it c}-axis, it is only displaced about 0.01\AA.  If the displaced
atoms also have displacement components along the a- or b-axes (which average
to zero) the total local displacements can be much larger.  Similarly if the
charge of the displaced atom is larger than {\it e} or only a fraction of a
charge, the displacement will be corresponding smaller or larger.

In this short paper we address the local ferroelectric distortions in TbMnO$_3$
and place upper limits on the magnitude of the atomic displacements from EXAFS
measurements. If there are moderately large displacements of some atoms but
only a small net component along the {\it c}-axis it will produce a series of
long and short bond lengths which will broaden the PDF functions and show up as
an amplitude change in the transmission EXAFS. We will show using difference
data that the changes in local distortions are very small and this rules out
significant displacements ($>$ 0.01\AA) of any atoms if the transition is
displacive.

\section{Experimental details and EXAFS data}
\label{exp}

A powdered sample and a small single crystal of TbMnO$_3$ were provided by
Kimura.  The powdered sample of TbMnO$_3$ was prepared by the solid state
reaction.  Powders of Tb$_4$O$_7$ and Mn$_2$O$_3$ were weighted to the
prescribed ratios, mixed, and well ground.  The mixture was heated at 1200
$^{\circ}$C in air for 16 hours.  After grinding, it was reheated at 1350
$^{\circ}$C for 24 hours.  A measurement of the powder x-ray diffraction
revealed that the resultant powder is of single phase with the $Pbnm$
orthorhombic structure at room temperature.  A single crystal of TbMnO3 was
grown by the floating zone technique, as previously described.\cite{Kimura05}
The crystal was oriented using Laue x-ray diffraction and patterns, and cut
into thin plates with the widest faces perpendicular to the crystallographic
principal axes. 

EXAFS data were collected in 2005 (Run1) at the Mn K- and Tb L$_3$-edges for a
powdered sample in transmission mode on Beamline 10-2 at SSRL using a Si 111
monochromator in focused mode; the slit height was 0.7 mm and the energy
resolution was about 2 eV. For these data, changes in the spectra with T were
very small as T increased through the ferroelectric transition temperature
T$_c$. Some polarized EXAFS (along the {\it c}-axis and perpendicular to the
{\it c}-axis) were later collected at NSLS in fluorescence mode but the
signal-to-noise was lower (in part due to small Bragg peaks from the sample),
and limits on changes in amplitude were poorer than for the powder transmission
data. To improve the transmission data and check reproducibility, a second
transmission data set was collected at SSRL in 2006 (Run 2) on Beamline 10-2
using the same monochromator. Again there was no obvious change in the EXAFS
plots as T increased through T$_c$. Note that the powder data are sensitive to
displacements in all directions.

\begin{figure}
\includegraphics[width=3.2in]{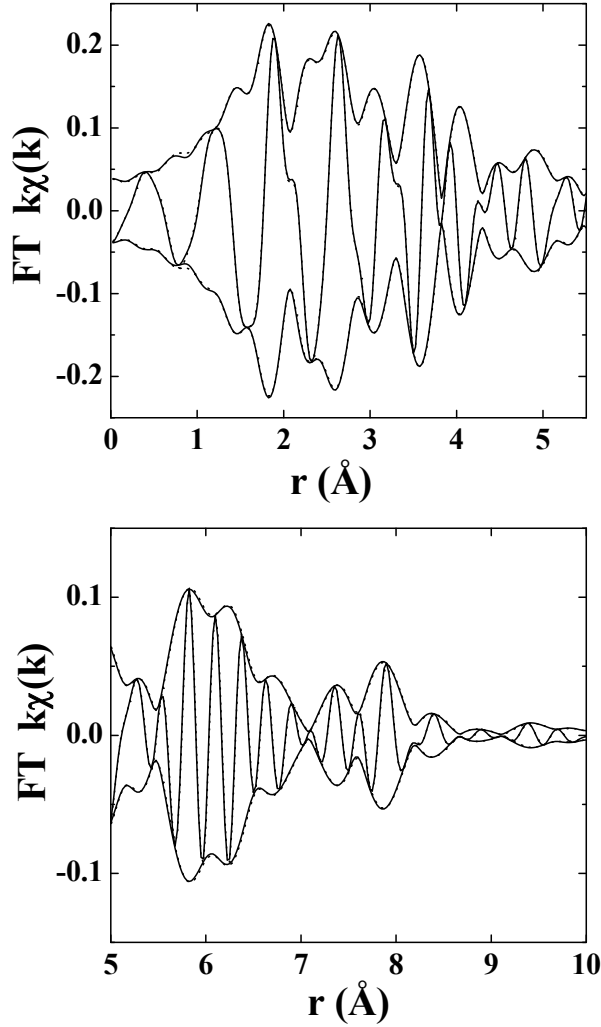}

\caption{Tb L$_3$-edge Fourier Transform data for TbMnO$_3$; k-range 3.5-12.8
\AA$^{-1}$ with a Gaussian rounding of 0.3\AA$^{-1}$. Solid line - Run 2, dotted
line - Run 1. Top is from 0-5.5 \AA, bottom from 5-10\AA{} (note different
y-scale). Even for the higher range in r the data overlap so well that it is
difficult to see both lines. This shows the high reproducibility of the data.
The fast oscillation in this figure and in subsequent r-space plots, is the
real part of the FT , R, while the envelops at the top and bottom are $\pm$
$\sqrt{I^2 + R^2}$ where I is the imaginary part of the FT.}

\label{tb_data}
\end{figure}

To improve the signal-to noise we next averaged the data in three groups for
each run: Group G1 - temperatures below T$_c$ (27K), G2 - between the Neal
temperature T$_N$ (42K) and T$_c$, and G3 - above T$_N$, between 42 and 50K;
the number of files averaged was generally between 3 and 6. In Fig
\ref{tb_data} we show the averaged Tb Fourier Transform (FT) data (L$_3$ edge)
at low T$<$ 27K, collected in runs 1 and 2. In this figure the reproducibility
of the data is so good that the difference between the two data sets is not
visible all the way to 10\AA. On this plot several Tb-O peaks occur from
1.8-2.0\AA, Tb-Mn peaks occur from 2.5-3.0\AA, and the nearest Tb-Tb peaks
occur near 3.5\AA {} (note these are peak positions in the EXAFS spectra which
have a shift from the actual bond length).  Above about 3.5\AA {} there is
increasing overlap between Tb-O, Tb-Mn and Tb-Tb and the peaks cannot be
assigned to specific pairs without a detailed fit.

\begin{figure}
\includegraphics[width=3.0in]{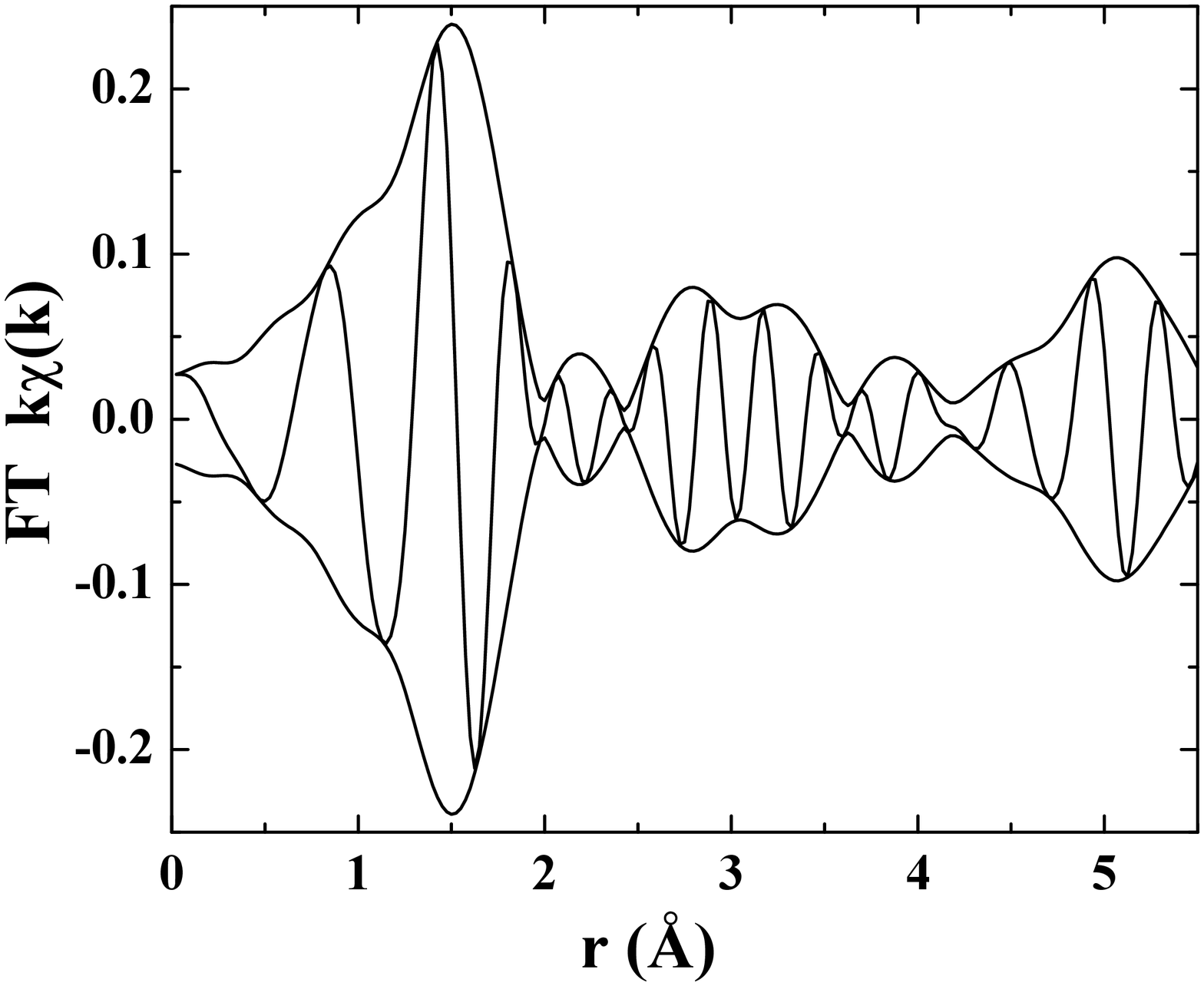}

\caption{The Mn K-edge r-space data for Run2; the data for run 1 are similar.
The FT range is 4-12.5 \AA$^{-1}$ with a Gaussian rounding of 0.3\AA$^{-1}$. }

\label{mn_data}
\end{figure}

In Fig \ref{mn_data} we show the corresponding averaged data from run 2 at low
T (T$<$ 27K) for the Mn K-edge. Here the Mn-O peaks occur near 1.5\AA{} and
the Mn-Tb peaks from 2.7-3\AA; the first Mn-Mn peaks occur near 3.4 and 3.6\AA.
Again, above 3.5\AA {} the peaks are a mixture of Mn-O, Mn-Tb and Mn-Mn and not
easily assigned to specific atom-pairs.


\section{Analysis}
\label{analysis}

\begin{figure}
\includegraphics[width=3.0in]{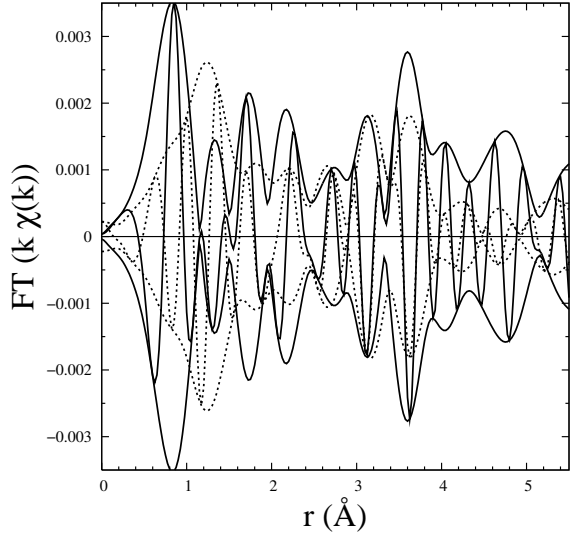}

\caption{The difference data G2-G1 at the Tb L$_3$-edge; solid line - Run 1,
dotted lines - Run 2. The difference for both runs is about 1\% of the magnitude
shown in Fig \ref{tb_data}. }

\label{tb_sub_mid_low} 
\end{figure}

To explore the magnitude of any ferroelectrically induced local structure
changes as the temperature increases through T$_c$, we have subtracted the
averaged data G1 (below T$_c$) from the averaged data G2 (between T$_c$ and
T$_N$). In Fig.  \ref{tb_sub_mid_low} we plot and compare this difference at
the Tb L$_3$ edge for both experimental runs.  The difference is very small for
both cases and is $\sim$ 1\% of the magnitude of the original data - compare
y-scales on Figs. \ref{tb_sub_mid_low} and \ref{tb_data}. [Note that
differences below r $\sim$  1.0\AA{} are not real as this is where small errors
in determining the background show up.] The small magnitude of the difference
shows the high reproducibility of the data and indicates that displacements of
Mn or O relative to a Tb atom are very small. A further requirement for a real
distortion is that the phase of the real part R of the Fourier Transform
difference data for the two experimental runs must be the same - over much of
the r-range they are not. Only in a few places does the phase for the two runs
agree (near 3\AA{}  - the Tb-Mn peaks, and near 2\AA{} - the Tb-O peaks). These
would be the only places where the difference data might suggest a real lattice
distortion. 

\begin{figure}
\includegraphics[width=3.0in]{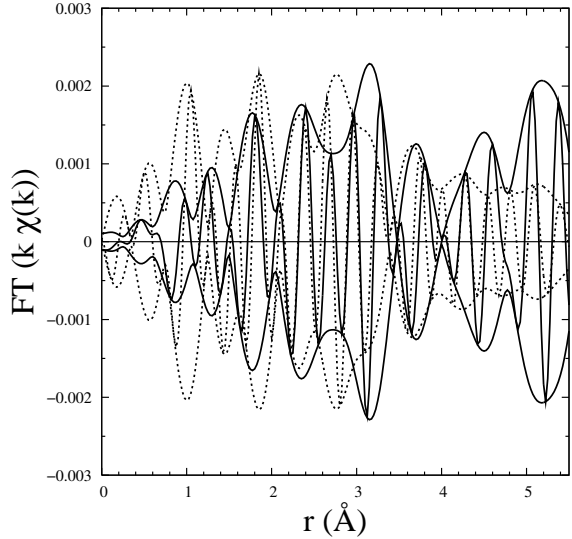}
\caption{The difference data G2-G1 at the Mn K-edge; solid line Run 1, dotted
lines Run 2. Here the difference is slightly smaller than for Tb but again about
1-2\% of the original data - Fig. \ref{mn_data}. } 

\label{mn_sub_mid_low}
\end{figure}

A similar result is obtained for the Mn K-edge difference data plotted in Fig.
\ref{mn_sub_mid_low}. Again the difference is comparable for both runs and is
about 1-2\% of the magnitude of the original data indicating at most very tiny
changes in the local distortions about the Mn atom as the sample becomes
ferroelectric. Again the phase of the real part R of the transform varies for
the two runs - only near 3\AA{} - the Mn-Tb peaks, does the phase suggest a
possible real effect. For both edges the difference is near the reproducibility
level, and smaller differences would be difficult to obtain. From a
consideration of the phases there may be a small change in the Tb-Mn, Mn-Tb
distances or in the Tb-O distances.

Another comparison can be made across the Neal temperature T$_N$ - if the
magnetic coupling causes a distortion of the Mn-O-Mn buckling angle that
depends on whether the Mn spins locally are parallel or antiparallel, similar
to that proposed by Aliouane {\it et al.} when a field is applied along the
{\it b}-axis,\cite{Aliouane06} a small off-center O distortion might then first
appear at T$_N$ but be randomly oriented for T$>$ T$_c$; however it would
produce longer and shorter bonds and show up in the EXAFS data at T$_N$. In
that case there would be no change in the local distortions across T$_c$ (same
numbers of long and short bonds); instead there would be an ordering of the
off-center displacements.  To look for changes in the local distortions at
T$_N$, we calculated difference data G3-G2 (above and below  T$_N$) for both
edges and for both experimental runs; the calculations  yield plots with
slightly smaller amplitudes than in Figs.  \ref{tb_sub_mid_low} or
\ref{mn_sub_mid_low} and are not plotted.  These results show no evidence for a
significant distortion appearing at T$_N$.  

\begin{figure}
\includegraphics[width=3.0in]{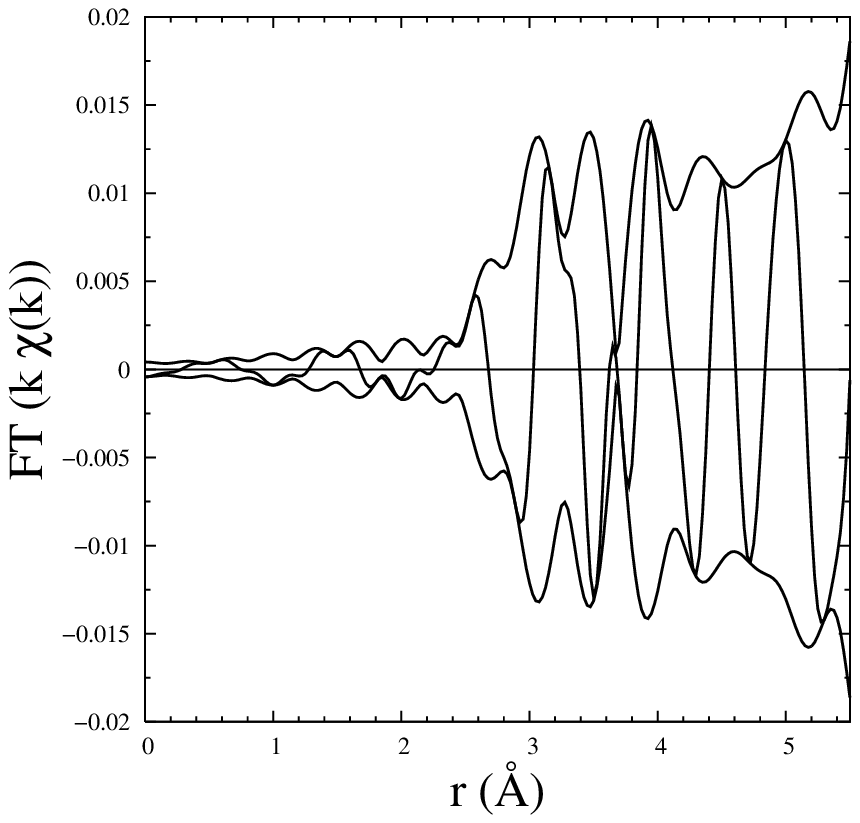}

\caption{The theoretical difference plots for a displacement of Tb by 0.01\AA{}
along the {\it c}-axis, relative to the rest of the unit cell.  The difference
is much larger than observed experimentally for the metal-metal atoms pairs
(above 2.8\AA) but comparable to the amplitude for r close to the Tb-O
distances.  }

\label{tb_c-axis-diff} 
\end{figure} 

\begin{figure}[t]
\vspace{0.2in}
\includegraphics[width=1.0in]{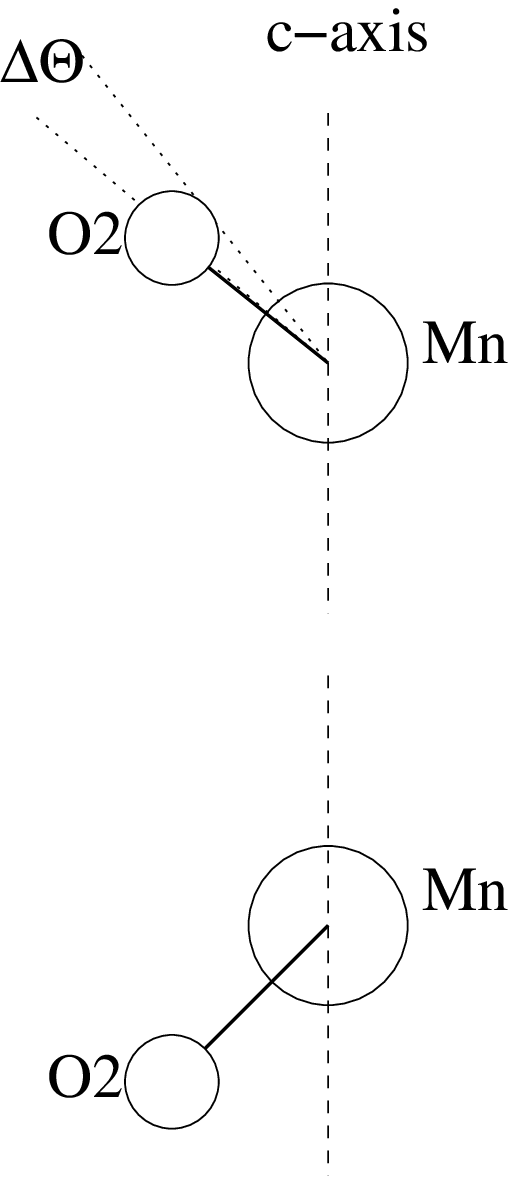}

\caption{A sketch of the Mn-O2-Mn linkage viewed along the Mn-O-Mn chains (the
{\it x}-axis in a pseudo-cubic perovskite unit cell) - the second Mn atom is
behind the first one. The O2 atoms are displaced above or below the {\it xy}
plane such that the angle between the {\it c}-axis and the plane containing the
Mn-O-Mn linkage is about 55$^{\circ}$.  A rotation of the plane of the Mn-O2-Mn
linkage that produces a positive {\it c}-axis displacement ($\Delta \theta$ in
figure) generates both positive and negative {\it a,b}-axis displacements which
would lead to zero polarization along the {\it a}- or {\it b}-axes.  }

\label{linkage}
\end{figure}

The above plots show that across T$_c$ there is very little change in the EXAFS
spectra - to place limits on the possible distortions we need to calculate how
much the spectra should change for various small displacements of different
neighbors atoms.  In Fig. \ref{tb_c-axis-diff} we plot the theoretical
difference for a displacement of the Tb atoms by 0.01\AA{} relative to the rest
of the unit cell. Here we have used $\sigma^2$ = 0.0025\AA$^2$ to include
thermal broadening (zero-point-motion) effects. The difference data do not
change much for increases in $\sigma^2$ for this edge or for the Mn edge
discussed below. Comparing this plot with Fig.  \ref{tb_sub_mid_low}
immediately gives the result that any displacements of Tb relative to Mn/O must
be much smaller than 0.01\AA {} (likely $<$ 0.005\AA). For the Tb edge, a {\it
c}-axis displacement of the O1 atoms would not change the Tb-O1 bond length
significantly. However unlike the Mn K-edge data discussed below, the Tb-O2
distribution is sensitive to displacements of Tb along the {\it c}-axis
relative to the unit cell - or conversely {\it c}-axis displacements of the O2
in the unit cell.  In addition it should be noted that the Mn-O2-Mn linkage is
not in the {\it ab}-plane but lies in a plane that is roughly 55$^{\circ}$ from
the {\it c}-axis. If this plane rotates slightly, the O2 moves towards one Tb
atom and away from another; them the projection in the {\it ab}-plane produces a
buckling of the (in-plane) Mn-O-Mn group that increases or decreases depending
on the sign of the rotation angle $\delta \theta$.  (See Fig. \ref{linkage})
Such rotations (equal numbers of $\pm$ $\delta \theta$) would not have any net
polarization along the {\it a} and{\it b} axes but would produce a net O2
displacement along the {\it c}-axis.  A 0.01\AA{} O2 (rotation) displacement of
this type was also tried and gave a similar magnitude in the difference
calculation near 2.2\AA{} as obtained by displacing the Tb relative to the unit
cell; however the phase does not agree well with the experimental data
suggesting a 0.01\AA{} displacement is too large.


\begin{figure}

\includegraphics[width=3.0in]{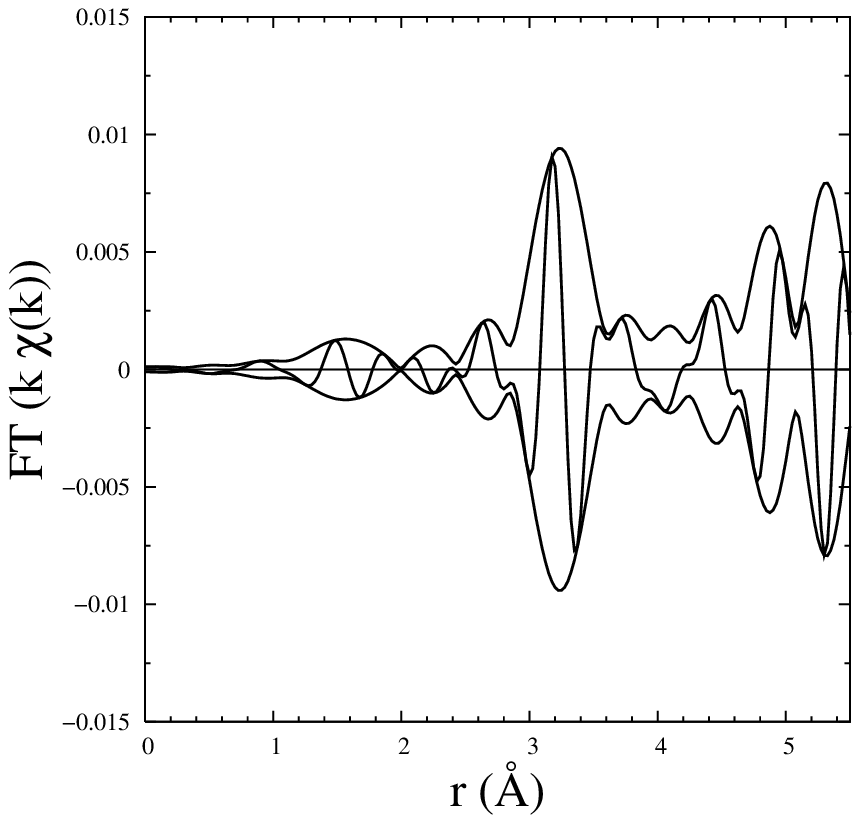}
\caption{The theoretical difference plots for a displacement of Mn by 0.01\AA{}
along the {\it c}-axis, relative to the rest of the unit cell.  The difference
is much larger than observed experimentally for the metal-metal atoms pairs
(above 2.8\AA) but comparable  to the amplitude for {\it r} close to the Tb-O
distances.  }

\label{mn_c-axis-diff} 
\end{figure} 

In Fig. \ref{mn_c-axis-diff} we plot the corresponding theoretical difference
for a Mn atom displacement of 0.01\AA{} relative to the rest of the unit cell;
again we have used $\sigma^2$ = 0.0025\AA$^2$.  Comparing this plot with Fig.
\ref{mn_sub_mid_low} immediately shows that any displacements of Mn relative to
Tb/O must also be much smaller than 0.01\AA.  For the Mn-O neighbors The main
Mn-O peak is at 1.5\AA; in the theoretical difference data for shifted Mn
atoms, the amplitude is comparable to that in the experimental difference data
- so a 0.01\AA {} displacement of O relative to Mn is not ruled out from the
  magnitude of the Mn K-edge difference results.  However the shape and phase
differ considerable between the two runs suggesting a much smaller change of
Mn-O bonds than 0.01\AA. Note, for Mn-O, only the O1 atoms along the {\it
c}-axis produce a significant change for a {\it c}-axis displacement of Mn -
the Mn-O distances for O2 atoms (which are roughly in the {\it ab} plane)
change little for a {\it c}-axis displacement of Mn.

Finally we return to the magnitude of the dipole moment per unit cell inferred
from the polarization and consider it is caused by a {\it c}-axis displacement
of only the O2 atoms. In the unit cell there are four formula units which means
eight O2 atoms (charge -2) per cell. If all are displaced the same amount, then
the net dipole moment for the cell is 8$\times$(2{\it e})$\times$$\delta$r =
0.01 {\it e}\AA.  This would infer a tiny displacement of  $\delta$r $\sim$ 0.6
$\times$10$^{-3}$\AA{} which is well below the limits obtained from the above
measurements.  If not all the O2 atoms are displaced or the effective charge on
the O2 ion is less than 2, then the displacement will be correspondingly
larger. A similar estimate is obtained by displacing the metal atoms along the
{\it c}-axis, relative to the O atoms.

In summary we show from difference EXAFS data that displacements of atoms must
be at or below 0.005\AA{} which rules out larger displacements which have {\it
a} and {\it b} components that average to zero. Assuming only O2 atom
displacements along the {\it c}-axis and a full, -2{\it e}  charge for this O
atom, the needed displacement must be at least 6$\times$10$^{-4}$\AA; these
results constrain $\delta r$ - 6$\times$10$^{-4}$\AA{} $<$ $\delta r$ $<$
5$\times$10$^{-3}$\AA. At this level one must also consider electronic
contributions arising from changes in the distributions of charge on the
various atoms. 

%
%

\acknowledgments{The work at UCSC was partially supported by NSF grant
DMR0301971.   Work at Los Alamos was performed under the auspices of the US
Department of Energy (DOE).The transmission EXAFS experiments were performed at
Stanford Synchrotron Radiation laboratory (beamline 10-2),  which is operated
by the DOE, Division of Chemical Sciences, and by the NIH, Biomedical Resource
Technology Program, Division of Research Resources. The polarized EXAFS were
carried out at the National Synchrotron Light Source, Brookhaven National
Laboratory (beamline 19A), which was supported by the DOE, Office of Science,
Office of Basic Energy Science, under contract No.  DE-AC02-98CH10886.}

%
%

\bibliographystyle{apsrev}

\bibliography{/home/users/bridges/bib/bibli}

\end{document}